\documentclass[letterpaper]{jpconf}
\usepackage{graphicx}

\begin{document}
\title{Status of the PICASSO experiment for spin-dependent Dark Matter searches}

\author{Marie-C\'{e}cile Piro (on behalf of the PICASSO collaboration)}

\address{Department of Particle Physics}
\address{University of Montreal}
\ead{piro@lps.umontreal.ca}

\begin{abstract}
The PICASSO project is using superheated droplets of C$_4$F$_{10}$ for the direct detection of Dark Matter candidates in the {\it spin-dependent} (SD) sector. The total setup includes 32 detectors installed in the SNOLAB underground laboratory in Sudbury (Ontario, Canada). With a concentrated effort in detector purification and with new discrimination tools now available for analysis, Picasso published competitive results in June 2009 \cite{publi2009} and became the leading experiment in the SD sector of direct dark matter searches. The present level of sensitivity is at 0.16 pb on protons at 90\% C.L. (M$_W$= 24GeV/c$^2$) following an analysis of two detectors only. The rest of the detectors are now in the process of being analyzed and the experimental search continues in order to further improve the limits or hopefully discover a signal of dark matter. The status of the experiment and the ongoing analysis will be presented.
\end{abstract}

\section{Introduction}
The astronomical and cosmological observations strongly suggest the presence of Dark Matter and show that only 1\% of the matter of the Universe is luminous and 23\% of all the matter should be of a new exotic kind \cite{bennett} : Cold Dark Matter (CDM). The preferred candidate for particle physicists is the neutralino, the Lightest Supersymmetric Particle (LSP). It is a very Weakly Interacting Massive Particle (WIMP) and is a natural candidate in the Minimal SuperSymmetric Model (MSSM) \cite{supersym}. PICASSO (Project In CAnada to Search for Supersymmetric Objects) is one of the many worldwide efforts to hunt for dark matter through the direct detection of neutralinos via their SD interactions with $^{19}$F nuclei \cite{ellis91,divari,bednyako}. The goal of the PICASSO project at SNOLAB is to exploit the favorable properties of $^{19}$F by using the superheated droplet detection technique, which is based on the operation principle of the classic bubble chamber \cite{glaser52,NC94}. 

\section{Detector principle and Response}
\begin{figure}[htb]
\begin{center}
\includegraphics[width=.65\textwidth]{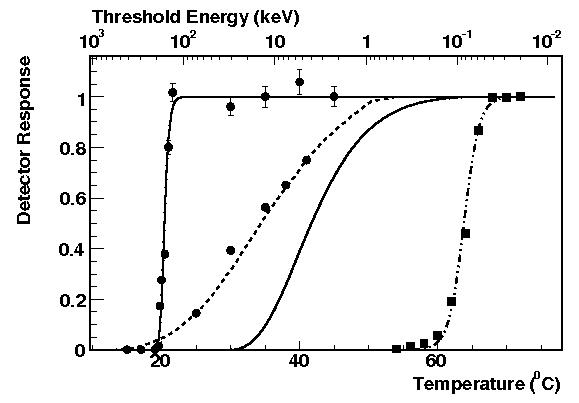}
\end{center}
\caption{Detector response to different types of particles as a function of temperature for 
detectors loaded with C$_4$F$_{10}$ droplets of $\sim$200$\mu$m in diameter. From left to right: 
alpha particles of 5.6 MeV in a detector spiked with $^{226}$Ra (fit to data points represented 
by continuous line);  nuclear recoils from fast neutrons of an AmBe source compared to 
simulations (dotted line); expected response for nuclear recoils following scattering of 
a 50 GeVc$^{-2}$ WIMP (continuous line); response to 1.275 MeV gamma rays of a $^{22}$Na 
source (dashed line).}\label{Fig:1}
\end{figure}

A PICASSO detector is a 4.5L cylindrical acrylic module, closed on top by a stainless steel lid sealed with polyurethane O-rings. Thirty-two of these modules are presently installed at SNOLAB. The active detector material consists of droplets of C$_4$F$_{10}$, a liquid which is at ambient temperature and pressure in a metastable, superheated state. The total mass of C$_4$F$_{10}$ amounts to 85g per detector module. When an ionizing particle deposits enough energy in a droplet, it creates a heat spike that induces a phase transition, changing the droplet into a vapor bubble.  This phase transition is accompanied by an acoustic signal which is recorded by nine piezoelectric-electric transducers. Calibrations with mono energetic neutron beams allow us to know precisely the energy thresholds of the detector as a function of pressure and temperature \cite{NIM}. This is important in order to predict the response curve for WIMPs, since neutron induced nuclear recoils are similar to those of WIMPs. The calibration runs are done at the Tandem accelerator facility at University of Montreal. The use of the light target nucleus $^{19}$F, together with a low detection threshold of 2 keV for recoil nuclei, makes PICASSO particularly sensitive to low-mass WIMPS. This is important as few experiments are sensitive below 15 GeVc$^{-2}$ and it is precisely in this mass range where recent interpretations of the DAMA/LIBRA annual modulations in terms of SD interactions remain partially unchecked \cite{bernabei,savage}. The detector response to different types of particles as a function of temperature is shown in Fig.\ref{Fig:1}. 

The different detector responses as a function of temperature allow us to discriminate a possible neutralino signal from backgrounds when we expose the detectors in the mine for counting. The operating temperature in each sub-unit can be varied independently from 20$^{\circ}$C to 55$^{\circ}$C and the temperature is regulated with a precision of  $\pm$ 0.1$^{\circ}$C. In this range of temperatures, the detectors are  mainly sensitive to alpha-particles and neutrons. In order to reduce these two kinds of background, the PICASSO detectors are carefully purified and the set-up is installed at 2 km underground at SNOLAB.

\section{Detector fabrication and purification}
Each PICASSO detector contains an emulsion of superheated droplets dispersed in a polymerized gel. The C$_4$F$_{10}$ droplets are dispersed in a gel of polyacrylamide forming a smooth interface. Auxiliary chemicals are added to control and improve the polymerization process. The emulsion of the droplets is created by a magnetic stirrer which is controlled such as to obtain a droplet size distribution centered around 200 $\mu$m in diameter. To create a homogeneous and uniform distribution of the droplets in the detector, a solution of cesium chloride salt (CsCl) is added to the matrix in order to equalize the densities of the C$_4$F$_{10}$ droplets and the gel solution at a value of 1.6g/cm$^3$. Since C$_4$F$_{10}$ has a boiling point of -1.7$^{\circ}$C, the fabrication process for the suspension takes place at -20$^{\circ}$C. In order to reduce the alpha-particle induced background, the detector material has to be purified. Purification is performed using Hydrous Zirconium Oxide (HZrO), a precipitate that attracts actinides like U, Th. The entire fabrication process is done in a class 10000 clean room.

\section{Analysis and discrimination tools}

\begin{figure}[htb]
\begin{center}
\includegraphics[width=.55\textwidth]{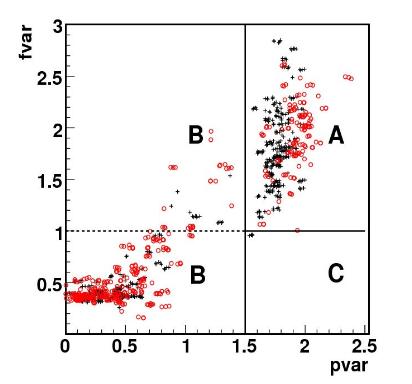}
\end{center}
\caption{The signal energy and frequency related variables  {\it pVar} and {\it fVar} allow the classification of events into distinct categories at 35$^{\circ}$C. Shown are data from n- calibration (black crosses) and WIMP runs (red circles). Neutron induced nuclear recoils from AmBe calibrations are located in region A; alpha particle events from WIMP runs appear displaced to the right in region A; electronic and acoustic noise populate the lower region B; fracture events produced by primary events in the polymer would be located in region C, but do not occur in this detector.}\label{Fig:2}
\end{figure}

A search by PICASSO for signatures in the recorded transducer signals showed that the intensity and frequency content of the acoustic signals contains information about the nature of the primary event \cite{aubin08}. From this information, variables can be constructed, which allow an event by event discrimination between particle and non-particle induced backgrounds. The first variable is named {\it pVar} and is based on the signal energy; it depends on the temperature and allows us to discriminate neutron induced recoils from backgrounds. The second variable is the frequency variable called {\it fVar} and is based on the power spectrum of the fast Fourier Transform (FFT) of the signal. These two parameters allow to separate particle-events due to WIMPS, alphas and neutrons, from the non-particle-induced events, like fractures and mechanical noises. Combining these two discrimination tools, we can determine a window to localize the nuclear recoil events expected for the interaction of neutralino induced events with the active liquid (Fig.\ref{Fig:2}).

\section{Experimental set-up and recent results obtained}
\begin{figure}[htb]
\begin{center}
\includegraphics[width=.55\textwidth]{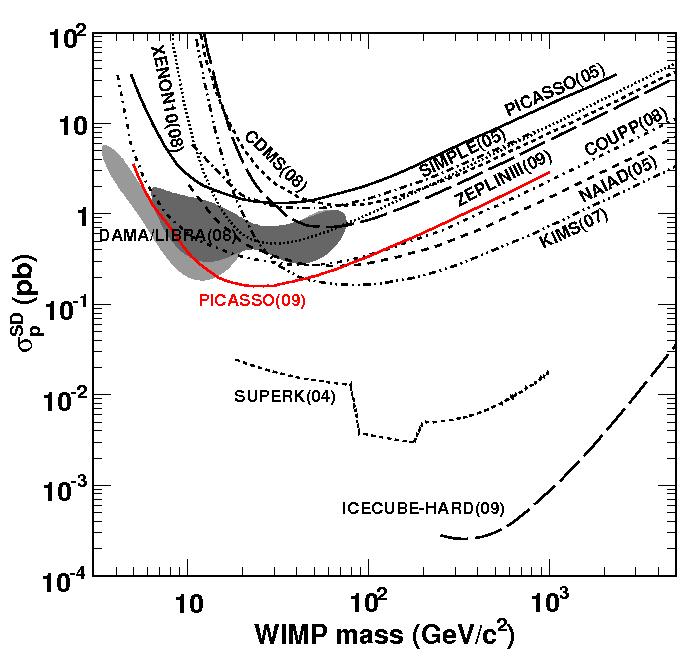}
\end{center}
\caption{Exclusion curve obtained with the two detectors only (Det. 71 and 72). \cite{publi2009}}\label{Fig:3}
\end{figure}

The experimental set-up is installed 2070 meters underground at SNOLAB at the Creighton mine in Sudbury. At this depth, the neutron flux coming from the ambient cosmic muon flux is largely reduced, with an expected count rate at the level of 10$^{-5}$ counts h$^{-1}$ g$^{-1}$, which is two orders of magnitude below the count rate of the detectors which are the subject of this analysis. The set-up is now completed with all 32 detectors installed yielding a total active mass of 2.6 kg of C$_4$F$_{10}$. The new discrimination tools have been applied first to a set of two detectors (71 and 72) with a total target mass of $^{19}$F target mass of 65.06 $\pm$ 3.2 g and 69.0 $\pm$ 3.5 g, respectively. For WIMP masses around 24 GeV/c$^2$ and a total exposure of  13.75 $\pm$ 0.48 kgd, PICASSO obtained new limits on the SD cross section on $^{19}$F of $\sigma_F$  = 13.9 pb (90\% C.L.) which can be converted into cross section limits on protons and neutrons of $\sigma_p$ = 0.16 pb and $\sigma_n$ = 2.60 pb respectively at 90\% C.L. (Fig.\ref{Fig:3}). The analysis is presently being extended to the remaining 30 detectors and we expect that a very competitive limit will be obtained when the detectors have been fully characterized and the analysis has been completed.

\section{Next phase of the experiment and conclusion}
PICASSO recently developed a new generation of detectors using a new kind of gel without cesium chloride which is presently the main source of alpha-contamination. In this case, viscous glycerin and polyethylene glycol allow to suspend the C$_4$F$_{10}$ droplets without density matching. Many improvements have been brought to the various fabrication steps, in order to obtain the lowest possible alpha-background. The preliminary results are very encouraging and may lead to a background reduction of at least a factor 10. Presently, half of the set-up has been replaced by these new detectors and the full installation is expected to be completed before the end of the year. The analysis of the new generation of detectors is now in progress and studies of a complete discrimination between the particle induced recoils and the alpha-particles will allow PICASSO to improve substantially the level of its sensitivity in the spin-dependant sector.\\

\end{document}